\RequirePackage[2020-02-02]{latexrelease}
\documentclass[amssymb,prb,twocolumn,showpacs]{revtex4}
\usepackage{epsfig}
\usepackage{dcolumn}
\usepackage{amsmath}
\begin{document}
\title{\bf\ Coherent states and their superpositions (cat states) in microwave-induced resistance oscillations.}
\author{J. I\~narrea$^{1,2}$ }

\address{$^1$Escuela Polit\'ecnica
Superior,Universidad Carlos III,Leganes, Madrid, 28911, Spain\\
$^2$Unidad Asociada al Instituto de Ciencia de Materiales, CSIC,
Cantoblanco, Madrid, 28049, Spain.}

\begin{abstract}
We report a novel theoretical approach on the microwave-induced resistance oscillations based on
 the coherent states of the quantum harmonic oscillator. We first obtain an expression for
the coherent states of driven-quantum harmonic oscillators that are used,
in the  model of microwave-induced electron orbits, to calculate magnetoresistance under radiation.
Thus, we find that the principle of
minimum uncertainty of coherent states, involving time and energy, is at the heart of
photo-oscillations and zero resistance states.
Accordingly, we are able to explain important experimental evidence of this remarkable  effect.
Such as the physical origin of oscillations, their periodicity with the
inverse of the magnetic field, their peculiar  minima and maxima positions
 and the existence of zero resistance states.
We apply our theory to the
case of ultra-high mobility samples where  we appeal to the principle of quantum superposition of
coherent states and obtain that Schrodinger cat states (even and odd coherent states)
are key to explain magnetoresistance at these extreme mobilities.
 With them we explain the, experimentally obtained,
magnetoresistance resonance
peak shift to  a magnetic field where the cyclotron frequency equals half the
radiation frequency. This effect is similar to the one described in quantum optics
 as a second harmonic generation process. We also explain the magnetoresistance
 collapse, that take place  in the dark and with light. This effect is known
 as giant negative magnetoresistance. We generalize our results to
 study the case of a three-component or triangular Schrodinger cat state.

\end{abstract}
\maketitle
\section{Introduction}
The first idea of coherent states or quasiclassical states was introduced by Schrodinger\cite{sro} describing
minimum uncertainty constant-shape Gaussian wave packets of the quantum harmonic oscillator.
They were constructed by the quantum superposition of the stationary states of the harmonic oscillator. These
wave packets
\begin{figure}
\centering\epsfxsize=3.2in \epsfysize=3.1in
\epsffile{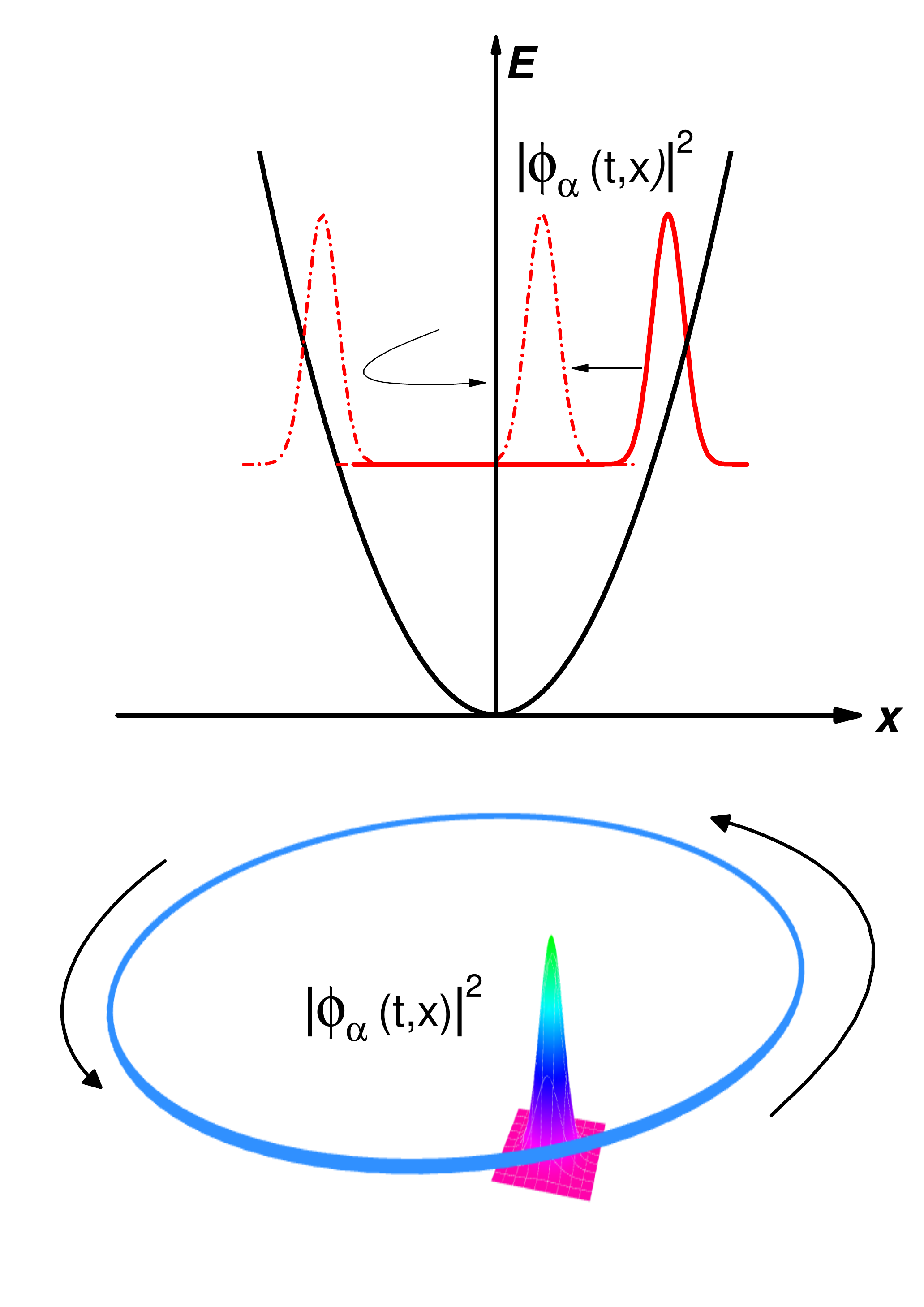}
\caption{Schematic diagrams of coherent state: The probability density of the coherent state is a
constant-shaped Gaussian
distribution, whose center oscillates in a harmonic oscillator potential similarly as
its classical counterpart.}
\end{figure}
displaced harmonically oscillating similarly as the classical counterpart\cite{sro}.
Later on, Glauber\cite{glau} applied the concept of coherent states to the
electromagnetic field being described by a sum of quantum field oscillators for
each field frequency or mode. These coherent states of electromagnetic radiation introduced by Glauber are
extensively used nowadays in quantum optics and form the foundation of another
even more exotic phenomena such as squeezed states\cite{dodonov} and Schrodinger cat states\cite{yurke,noel,dodonov2}.
Coherent states are an essential and powerful  tool in quantum mechanics when describing the dynamics of
quantum systems that are very close to a classical behaviour.
One remarkable example of this is the system of one electron under
the influence of a moderate and constant magnetic field ($B$).
The quantum mechanical solution of this problem lead us to
Landau states which are mere stationary states of a quantum harmonic oscillator.
Under low or moderate values of $B$, this system can be described by
an infinite superposition of Landau states, i.e., a coherent state of
the quantum harmonic oscillator. The resulting wave packet (wave function
probability density) oscillates classically at the cyclotron frequency ($w_{c}$)
inside the quadratic potential keeping constant the Gaussian shape (see Fig. 1)
and complying with the minimum uncertainty condition.

One important application example of coherent states in condensed matter
physics could be the remarkable phenomenon of microwave-induced resistance oscillations\cite{mani,zudov} (MIRO).
The reasons for MIRO to fit in a coherent state scenario is, first, they
are based on quantum states of harmonic oscillators, and second, the low or
moderate values of $B$ used in the experiments\cite{mani,zudov}.
MIRO   are one of most
striking radiation-matter interactions effects discovered in the last
two decades. They show up in the irradiated magnetoresistance ($R_{xx}$) of
high mobility two-dimensional electron systems (2DES) at low
temperature ($T\sim 1K$).
A pioneering  theoretical work on these systems  had  been already carried out by
 Ryzhii\cite{ryzhii} in the 70's.
 MIRO feature with a series of peculiar properties. They are periodic with
the inverse of $B$; the oscillations extrema show up at definite positions depending
on the frequency ratio $w/w_{c}$, $w$ being the radiation frequency;
and   when increasing radiation power,  MIRO minima do not naturally evolve under
$R_{xx}=0$, but the system keeps in a  vanishing resistance scenario, i.e.,
zero resistance states  (ZRS)\cite{mani,zudov}.
The discovery of MIRO  led to a great deal of
theoretical works back then. The three most important to date are:
the displacement model based on photo-assisted scattering from impurities or disorder
\cite{girvin}, the inelastic model\cite{dimi} based on the effect of radiation
on the electron distribution function and the microwave-driven
electron orbits model\cite{ina1,ina2}. According to the latter,
Landau states, under radiation, spatially and  harmonically  oscillate through the guiding center with the radiation frequency performing
 classical trajectories. In this swinging motion electrons
 are  scattered by charged impurities giving rise to irradiated magnetoresistance, i.e., MIRO.

In the present work we extend the microwave-driven electron orbit model introducing the
 coherent states of the quantum harmonic oscillator. Thus, we first obtain an
 expression for the wave function of the coherent states of  {\it radiation-driven} quantum harmonic oscillations
(driven Landau states or orbits).
With this wave function  we calculate the irradiated magnetoresistance with a
semiclassical Boltzmann transport theory where charged-impurity
scattering is considered. An important first result we obtain is that the time
it takes for a scattered electron to jump between Landau orbits or evolution time
between states, $\tau$, equals
the cyclotron period, $T_{c}=2\pi/w_{c}$. The rest of scattering processes
at different $\tau$ do not significantly contribute to the current. Interestingly
enough, this value for $\tau$ is hidden in the values that MIRO extrema
take in experiments. For instance, MIRO minima comply with $w/w_{c}=(j+1/4)$, $j$ being a positive integer.
Coherent states fulfill the minimum uncertainty condition and this decides which
Landau states can be reached by scattering. Thus, each pair of coherent
states connected by scattering have to fulfill the time-energy minimum uncertainty relation: $\Delta t \times \Delta E =h$ \cite{cohen}.
In our case, $\Delta t = \tau$ and then $2 \pi/w_{c} \times \Delta E= h$, implying that the
energy difference between coherent states involved in any scattering process must be, $\Delta E= \hbar w_{c}$ (see Fig. 2).
The rest of scattering processes, simply do not take place.
Under radiation due to the swinging nature of the driven-states,
the final minimum uncertainty state will be, at certain $B$, farther than in the dark yielding
MIRO peaks (see Fig. 3). In other $B$, they will be closer giving MIRO valleys.

 We further extend the
theory to explain the experimentally obtained\cite{zudov2,rui} resonance peak shift to $2w_{c}=w$ and
magnetoresistance collapse in samples with ultrahigh mobility. We are based on
one of the most fundamental principles of quantum mechanics: the superposition principle.
In our case the superposition of  coherent state to give rise to Schrodinger cat states
(even and odd coherent states). In those states, the Landau levels energy difference
 inside the coherent superposition is  $2 \hbar w_{c}$ with important consequences in 
 terms of scattering.
Accordingly, at sufficient radiation power the resonance peak now
rises at $2w_{c} =w$ instead of the expected position  $w_{c} =w$. 
Along with this resonance shift, Landau levels involved in scattering become totally
misaligned and due to the elastic nature of charged impurity scattering,
magnetoresistance dramatically drops off or plummets. The latter is observed both in the dark (giant
negative magnetoresistance) and under illumination. Therefore, we can consider
ultrahigh-mobility 2DES under low $B$ as a  platform for the
physical realization of Schrodinger cat states in condensed matter such as trapped-ions\cite{matos}
and Bose-Einstein condensates\cite{cirac}
The most interesting application example of this novel platform
would be in the implementation of qubits for
quantum computing.
We finally extent our model to three-component or triangular  Scrodinger cat states.

\begin{figure}
\centering\epsfxsize=3.2in \epsfysize=3.5in
\epsffile{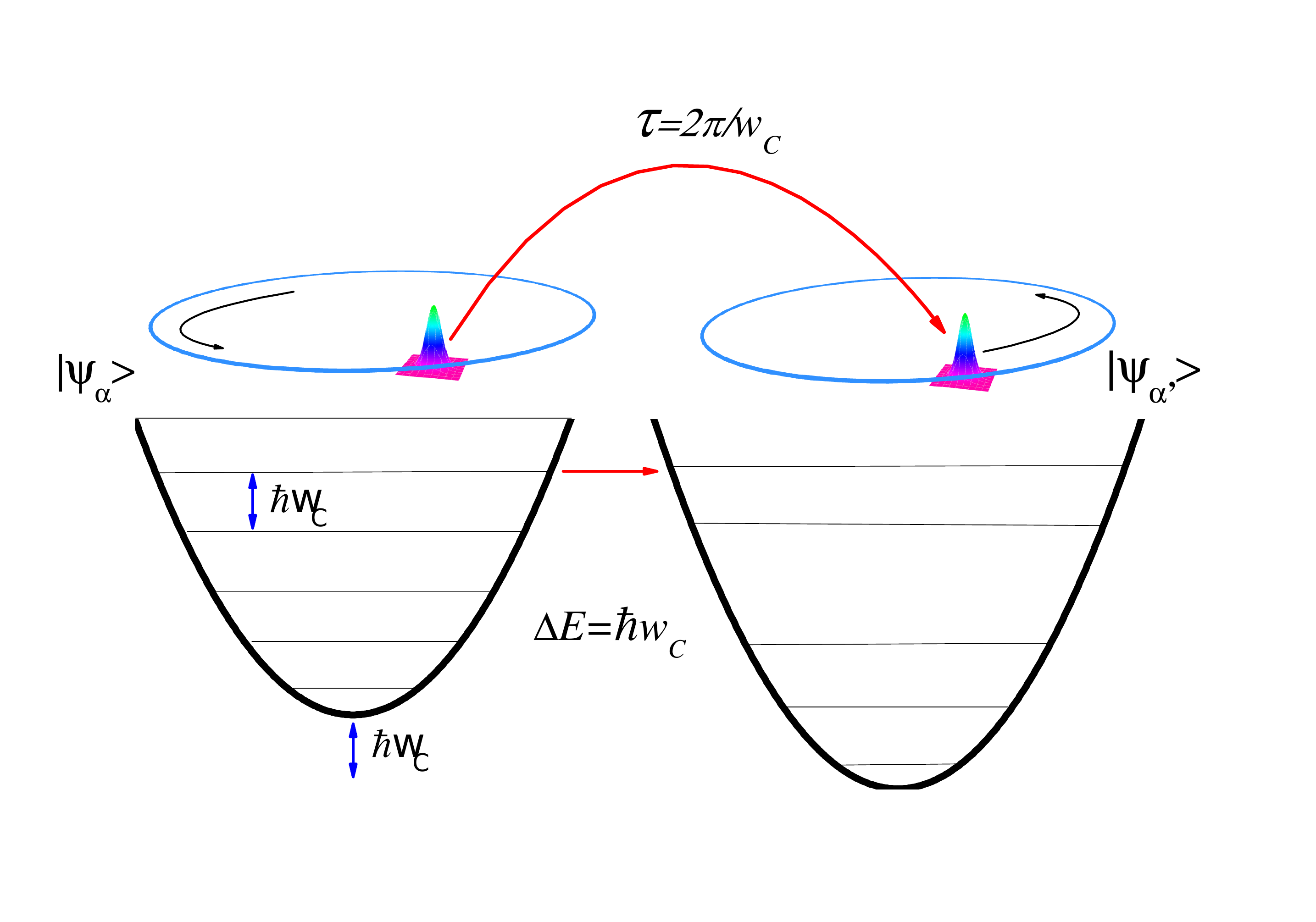}
\caption{Schematic diagram of scattering process between  coherent states $\Psi_{\alpha}$
and $\Psi_{\alpha'}$. The scattering is quasi-elastic because the scattering
source is based on charged impurities. The probability density for both
 coherent states is a
constant-shaped Gaussian wave packet and the process
evolution time is the cyclotron period. i.e., $\tau=2\pi /w_{c}=T_{c}$.
}
\end{figure}

\section{Theoretical model}
\subsection{Coherent states}
 We first obtain
an expression for the coherent states of radiation-driven
quantum harmonic oscillator  wave function. The starting point is the exact solution
of the time-dependent Sdchrodinger equation of a quantum harmonic oscillator under
a time-dependent force. In our problem this corresponds to
the electronic wave function for a 2DES in a
perpendicular $B$, a DC electric field $E_{DC}$ and MW radiation which is
considered semi-classically. The wave function is given by\cite{ina1,kerner,park}:
\begin{equation}
\Psi_{n}(x,t)=\phi_{n}(x-X(0)-x_{o}(t)) e^{-i w_{c} (n+1/2)t}\times e^{\frac{i}{\hbar}\Theta (t)}
\end{equation}
where,
\scriptsize
\begin{equation}
\Theta (t)=\left[m^{*}\frac{dx_{o}(t)}{dt}x-
\int_{0}^{t} {\it L} dt' \right] + X(0)\left[-m^{*}\frac{dx_{o}(t)}{dt}x+
\int_{0}^{t} {\it E_{0} \cos wt'} dt' \right]
\end{equation}
\normalsize
 $X(0)$ is the guiding center of the driven-Landau state,
$E_{0}$ the MW electric field intensity,
 $\phi_{n}$
is the solution for the Schr\"{o}dinger equation of the unforced
quantum harmonic oscillator (Landau states) and $x_{0}(t)$ is the classical
solution of a forced harmonic oscillator:
\begin{equation}
x_{0}=\frac{e E_{o}}{m^{*}\sqrt{(w_{c}^{2}-w^{2})^{2}+w^{2}\gamma^{2}}}\sin wt=A\sin wt
\end{equation}
where $\gamma$ is a phenomenologically-introduced damping factor
for the electronic interaction with acoustic phonons and
${L}$ is the classical Lagrangian.
Apart from phase factors, the wave function turns out to be
the same as a quantum harmonic oscillator where the center is
driven by $x_{cl}(t)$. Thus, all driven-Landau states harmonically oscillates   in phase
at the radiation frequency.

A coherent state denoted by $| \alpha \rangle$ is defined as the eigenvector of the anhilation operator $\hat{a}$
with eigenvalue $\alpha$
and can  be expressed as a superposition of quantum harmonic oscillator states\cite{cohen},
\begin{equation}
| \alpha \rangle=\sum_{n}c_{n}(\alpha)|\phi_{n}\rangle=e^{-|\alpha|^{2}/2}\sum_{n}\frac{\alpha^{n}}{\sqrt{n!}}|\phi_{n}\rangle
\end{equation}
The coherent state  $| \alpha \rangle$ can be also obtained  with the displacement operator $D(\alpha)$\cite{cohen}
acting on the quantum harmonic oscillator  ground state $|\phi_{0}\rangle$, $| \alpha \rangle= D(\alpha)|\phi_{0}\rangle$,
where the unitary operator $D(\alpha)$ is defined by: $D(\alpha)=e^{\alpha a^{\dagger}-\alpha^{\ast}a}$.
The coherent state in the position representation or wave function then reads like,
$\psi_{\alpha}(x)=\langle x| D(\alpha)|\phi_{0}\rangle$.

We observe, according to the obtained
MW-driven wave function  (Eq. 1),  that irrespective of
the MW oscillations, the Landau level structure remains unchanged with
respect to the dark situation; same Landau level index and energy. Then,
we conclude that the system is quantized, in the same way as the
unforced quantum harmonic oscillator.
 Thus, we can construct the coherent
states based on  driven-Landau states similarly as if they were in the dark\cite{cohen}:
\begin{eqnarray}
|\psi_{\alpha}(x,t) \rangle&=&e^{\frac{i}{\hbar}\Theta (t)} e^{-iw_{c}t/2} e^{-|\alpha|^{2}/2}\nonumber\\
&\times&\sum_{n}\frac{\alpha^{n}  e^{-inw_{c}t}}{\sqrt{n!}}|\phi_{n}(x-X(0)-x_{o}(t)       )\rangle\nonumber\\
\end{eqnarray}
Now applying the displacement operator on the MW-driven  ground state,
\begin{equation}
\psi_{\alpha}(x,t)=e^{\frac{i}{\hbar}\Theta (t)}\langle x| D(\alpha)|\phi_{0}(x-X(0)-x_{o}(t))\rangle
\end{equation}
 we can calculate the wave function corresponding to
the coherent state of the MW-driven quantum oscillator:
\large
\begin{widetext}
\begin{equation}
\Psi_{\alpha}(x,t)= e^{\frac{i}{\hbar}\Theta (t)} e^{i\vartheta_{\alpha}}e^{-iw_{c}t/2}e^{\frac{i}{\hbar}\langle p \rangle(t)x}
\phi_{0}[x-X(0)-x_{o}(t)-\langle x \rangle(t)]
\end{equation}
where,
\begin{equation}
\phi_{0}[x-X(0)-x_{o}(t)-\langle x \rangle(t)]=
\left (\frac{mw_{c}}{\pi\hbar}\right )^{1/4}
e^{-\left[\frac{x-X(0)-x_{o}(t)-\langle x \rangle(t)}{2\Delta x}\right]^{2}}
\end{equation}
\end{widetext}
\normalsize
$\langle x \rangle(t)$ and $\langle p  \rangle(t)$ are the position and
momentum mean values respectively\cite{cohen},
\begin{equation}
\langle x \rangle(t)=\sqrt{\frac{2\hbar}{m^{\ast} w_{c}}}|\alpha_{0}|\cos(w_{c}t-\varphi)
\end{equation}
 and
\begin{equation}
\langle p  \rangle(t)=-\sqrt{2m^{\ast}\hbar w_{c}}|\alpha_{0}|\sin(w_{c}t-\varphi)
\end{equation}
where we  have used that  $\alpha=|\alpha_{0}| e^{-(iw_{c}t-\varphi)}$.
$\Delta x$
is
the position uncertainty and the global phase facor, $ e^{i\vartheta_{\alpha}}=e^{\alpha^{\ast 2}-\alpha^{2}}$.
Then, the wave packet associated with $\Psi_{\alpha}(x,t)$ is therefore given by:
\begin{equation}
|\Psi_{\alpha}(x,t)|^{2}=|\phi_{0}[x-X(0)-x_{o}(t)-\langle x \rangle(t)]|^{2}
\end{equation}
As a result, this wave packet displaces under the influence of two harmonic motions,
one depending on the cyclotron frequency $w_{c}$ and the other on the radiation
frequency $w$.

If we want to calculate magnetoresistance, we first obtain the conductivity ${\sigma_{xx}}$ following
a semiclassical Boltzmann model\cite{ridley,ando,askerov},
 \begin{equation}
\sigma_{xx}=2e^{2} \int_{0}^{\infty} dE \rho_{i}(E) (\Delta X_{0})^{2}W_{I}\left( -\frac{df(E)}{dE}  \right)
\end{equation}
being $E$ the energy, $\rho_{i}(E)$ the density of
initial Landau states and $W_{I}$ is the scattering rate of
electrons with charged impurities.
We consider now that the scattering takes place between coherent states of quantum harmonic oscillators.
Thus, $\Delta X_{0}$ is the distance between the guiding centers  of the scattering-involved coherent states.

We first study  the dark case and
according to the Fermi's golden rule $W_{I}$ is given by,
\begin{equation}
  W_{I}=N_{i}\frac{2\pi}{\hbar}|<\psi_{\alpha^{'}}|V_{s}|\psi_{\alpha}>|^{2}\delta(E_{\alpha^{'}}-E_{\alpha})
\end{equation}
where $N_{i}$ is the total number of impurities, $\psi_{\alpha}$ and $\psi_{\alpha^{'}}$ are the wave functions  corresponding to the initial and final coherent states respectively,
 $V_{s}$ is the scattering potential for charged impurities\cite{ando}:
 $V_{s}=\sum_{q} V_{q}e^{i q_{x} x}= \sum_{q}\frac{e^{2}}{2 S \epsilon (q+q_{s})} e^{i q_{x} x}$,
$S$ being the sample surface, $\epsilon$ the dielectric
constant, $q_{TF}$ is the Thomas-Fermi screening
constant\cite{ando} and $q_{x}$ the $x$-component of $\overrightarrow{q}$,
the electron momentum change after the  scattering event.
 $E_{\alpha}$ and $E_{\alpha^{'}}$ stand for the coherent states initial and final energies respectively.
The $V_{s}$ matrix element is given by\cite{ridley,ando,askerov}:
\begin{equation}
|<\psi_{\alpha^{'}}|V_{r}|\psi_{\alpha}>|^{2}=\sum_{q}|V_{q}|^{2}|I_{\alpha,\alpha^{'}}|^{2}
\end{equation}
and the term $I_{\alpha,\alpha^{'}}$\cite{ridley,ando,askerov},
\begin{widetext}
\begin{equation}
I_{\alpha,\alpha^{'}}=\int^{\infty}_{-\infty} \psi_{\alpha^{'}}(x-X^{'}(0)-\langle x' \rangle(t')) e^{i q_{x} x}\psi_{\alpha}(x-X(0)-\langle x \rangle(t)) dx
\end{equation}
\end{widetext}
where,
\begin{equation}
\psi_{\alpha}=e^{i\vartheta_{\alpha}}e^{-iw_{c}t/2}e^{\frac{i}{\hbar}\langle p \rangle(t)x} \left (\frac{mw_{c}}{\pi\hbar}\right )^{1/4}
e^{-\left[\frac{x-X(0)-\langle x \rangle(t)}{2\Delta x}\right]^{2}}
\end{equation}
and similar expression for $\psi_{\alpha^{'}}$.
After lengthy algebra we obtain an expression for $|I_{\alpha,\alpha^{'}}|$,
\large
\begin{equation}
  I_{\alpha,\alpha^{'}}=e^{-\frac{[X^{'}(0)-X(0)+\langle x^{'} \rangle(t')-\langle x \rangle(t)]^{2}}{8(\Delta x)^{2}}}
  e^{-\frac{q_{x}^{2}(t)2(\Delta x)^{2}}{4}}
\end{equation}
\normalsize
where $q_{x}(t)$is given by,
\begin{equation}
q_{x}(t)=q_{x}+2\sqrt{2m\hbar w_{c}}[-|\alpha_{0}^{'}|\sin(w_{c}t') +|\alpha_{0}|\sin(w_{c}t)]
\end{equation}

On the other hand,
\begin{widetext}
\begin{equation}
\langle x^{'} \rangle(t^{'})-\langle x \rangle(t)=\sqrt{\frac{2\hbar}{m w_{c}}}\left[|\alpha_{0}^{'}|\cos(w_{c}t^{'}-\varphi)-\alpha_{0}|\cos(w_{c}t-\varphi)\right]
\simeq \sqrt{\frac{2\hbar}{m w_{c}}}|\alpha_{0}|2\sin(w_{c}(t+\frac{\tau}{2})-\varphi) \sin(-w_{c}\frac{\tau}{2})
\end{equation}
\end{widetext}
where $t$ and  $t^{'}$ are the initial and final times for the scattering event and $\tau$ is the evolution
time between coherent states. Thus, $t^{'}=t+\tau$. We have considered also that for the
experimental low values of $B$,   $|\alpha^{'}_{0}|\simeq|\alpha_{0}|$. Developing the square of
the above exponential we can finally get to,
\large
\begin{equation}
  I_{\alpha,\alpha^{'}}\propto e^{-2|\alpha_{0}|^{2}\sin^{2}(w_{c}(t+\frac{\tau}{2})-\varphi) \sin^{2}(w_{c}\frac{\tau}{2})}
\end{equation}
\normalsize
For experimental $B$,  $|\alpha_{0}|^{2} > 50$ and thus, $|I_{\alpha,\alpha^{'}}| \rightarrow 0$.
Accordingly, scattering rate and conductivity would be zero too. Nonetheless, there is an important exception when
$\tau$ equals the cyclotron time $T_{c}$:
\large
\begin{equation}
\tau=n \times \frac{2\pi}{w_{c}}
\end{equation}
\normalsize
$n$ being a positive integer.

\begin{figure}
\centering \epsfxsize=3.5in \epsfysize=3.5in
\epsffile{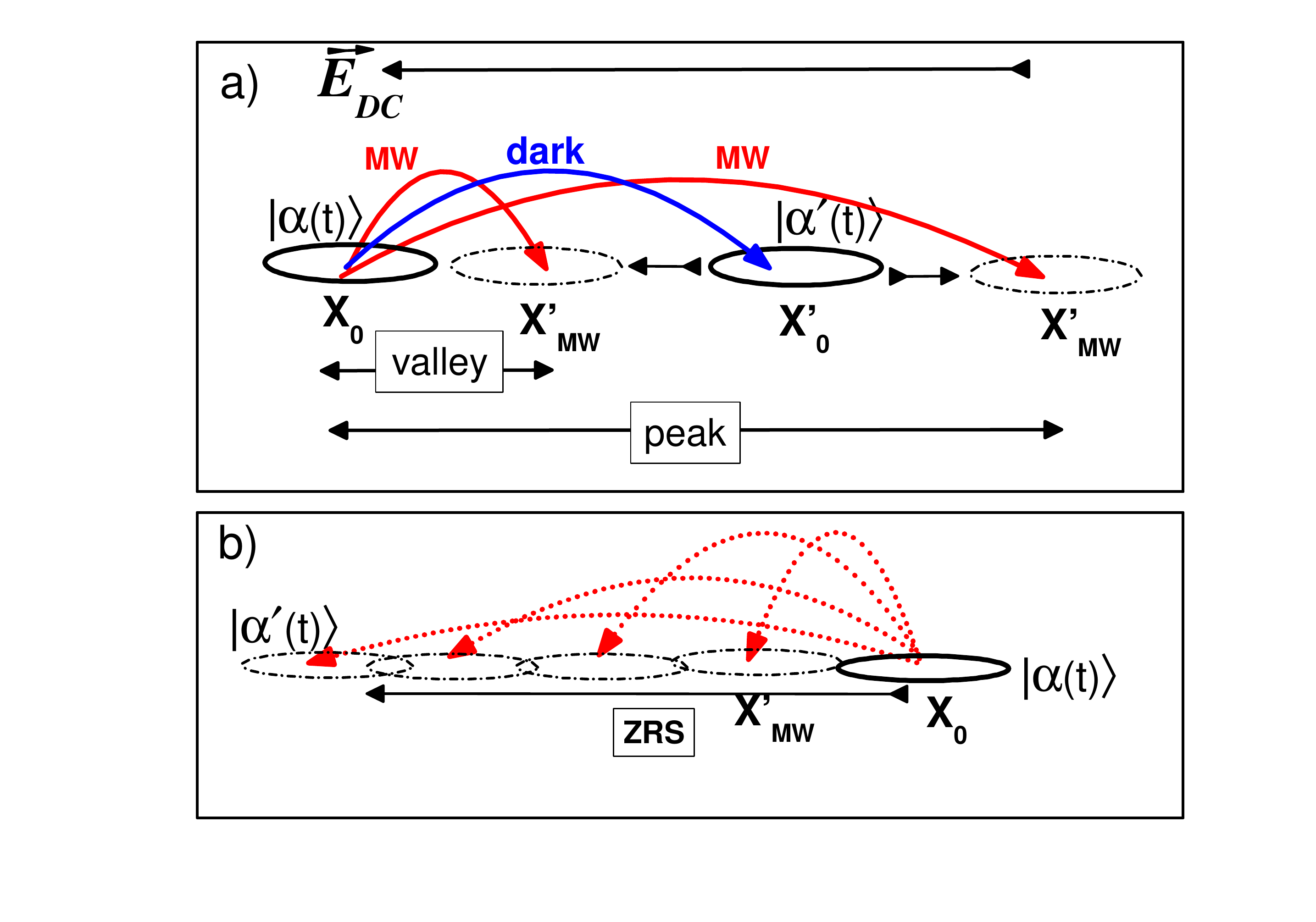}
\caption{Schematic diagrams for the different
situations regarding dark and light which include peaks, valleys and
zero resistance states. The final scattering state has to fulfilled
the minimum uncertainty principle.  a)  The final state in its oscillation
is further away from the initial state giving MIRO peaks. On the hand,
when is closer we obtain MIRO valleys.
 b)  Situation when MW power is high enough and the states go backwards.
In this scenario the final state ends up behind the initial state dark position and
the scattering jump can not take place. }
\end{figure}
 In other words, the scattered electron
begins and ends in the same position in the Landau orbit. Thus, only scattering processes fulfilling  the previous condition
of the evolution time will efficiently contribute to the current. The rest contributions can be neglected.
Finally the expression of  $|I_{\alpha,\alpha^{'}}|$ reads\cite{ridley},
\large
\begin{equation}
I_{\alpha,\alpha^{'}}= e^{-\left(\frac{(X^{'}(0)-X(0))^{2}}{8(\Delta x)^{2}}+\frac{q_{x}^{2}(\Delta x)^{2} }{2}\right)}
\end{equation}
\normalsize
This, in turn, lead us to a final expression for  $W_{I}$,
\begin{equation}
W_{I}=\frac{n_{i}e^{4}}{2 \pi \hbar \epsilon^{2}}\int \frac{e^{-q^{2}(\Delta x)^{2}}}{(q+q_{TF})^{2}}(1-\cos \theta)\delta(E_{\alpha^{'}}-E_{\alpha})d^{2}q
\end{equation}
where $n_{i}$ is the  charged impurity density,
 $\Delta X_{0}=X^{'}(0)-X(0)=[-q_{y}2(\Delta x)^{2}]$\cite{kerner} and $\theta$ is the scattering
angle.
The density of initial Landau states $\rho_{i}(E)$ can be obtained by
using the Poisson sum rules to get to\cite{ihn},
\begin{equation}
\rho_{i}(E)= \frac{m^{\ast}}{\pi \hbar^{2}}\left[1-2\cos \left(\frac{2\pi E}{\hbar w_{c}} e^{-\pi\Gamma/\hbar w_{c}} \right)\right]
\end{equation}
Gathering all terms and solving the energy integral,  we obtain an expression for $\sigma_{xx}$ that reads,
\begin{widetext}
\begin{equation}
\sigma_{xx}=\frac{n_{i}e^{6}m^{\ast}}{2\pi^{3}\hbar^{3}\epsilon^{2}}(\Delta X_{0})^{2}\frac{1}{\hbar w_{c}}\left(\frac{1+e^{-\pi\Gamma/\hbar w_{c}}}{1-e^{-\pi\Gamma/\hbar w_{c}}}\right)
\left(1-\frac{2\chi_{s}}{\sinh (\chi_{s}) } \cos\left(\frac{2\pi E_{F}}{\hbar w_{c}}\right) e^{-\pi\Gamma/\hbar w_{c}} \right) \int \frac{e^{-q^{2}(\Delta x)^{2}}}{(q+q_{TF})^{2}}(1-\cos \theta) d^{2}q
\end{equation}
\end{widetext}
where $\chi_{s}=2\pi^{2}k_{B}T/\hbar w_{c}$, $k_{B}$ being the Boltzmann constant and $E_{F}$ the Fermi energy.
To obtain
$R_{xx}$ we use the relation
$R_{xx}=\frac{\sigma_{xx}}{\sigma_{xx}^{2}+\sigma_{xy}^{2}}
\simeq\frac{\sigma_{xx}}{\sigma_{xy}^{2}}$, where
$\sigma_{xy}\simeq\frac{n_{e}e}{B}$ and
$\sigma_{xx}\ll\sigma_{xy}$, $n_{e}$ being the 2D electron density.

One important condition that features coherent states is that they minimize the Heisemberg uncertainty principle. Thus, for
the position-momentum uncertainties, $\Delta x_{\alpha} \Delta p_{\alpha}=\hbar/2$, where $\Delta x_{\alpha}$ and  $\Delta p_{\alpha}$ are given by,
\begin{eqnarray}
\Delta x_{\alpha}&=&\sqrt{\frac{\hbar}{2m^{\ast}w_{c}}} \\
\Delta p_{\alpha}&=&\sqrt{\frac{m^{\ast}\hbar w_{c}}{2}}
\end{eqnarray}
We can extend this to
the time-energy uncertainty relation and write\cite{cohen}, $\Delta t \Delta E=h$.
For our specific problem,  $\Delta t=\tau$ that implies  $\Delta E=\hbar w_{c}$, $\Delta E$ being
 the energy difference between scattering-involved coherent states. Thus, we obtain two conditions
for the scattering between coherent states to take place, first $\tau=\frac{2\pi}{w_{c}}$ and second,
the energy difference equals  $\hbar w_{c}$.
There are also physical reasons that endorse  the latter specially in high-mobility
samples where the levels are very narrow. In these systems the only efficient contributions to scattering are the ones corresponding
 to aligned Landau levels (see Fig. 2),
i.e.,
when $\Delta E=n \times \hbar w_{c}$. But the most intense  of them is when $n=1$
that corresponds to the closest in distance coherent state or smallest value
of  $\Delta X_{0}$, (see Eq. 22). And this agrees with  the point
that when $n=1$, the Heisemberg uncertainty is
minimized. The above scenario leads to the situation
where any misalignment will make the current collapse.

When we turn on the light, the term that is going to be mainly
affected in the $\sigma_{xx}$ expression is the distance between the guiding
center of the wave packets. i.e., $\Delta X_{0}$. This average distance now
turns into $\Delta X_{MW}$\cite{inarashba,inahole}
\begin{widetext}
\begin{eqnarray}
\Delta X_{MW}&=&X^{'}_{MW}-X_{MW}
=\left(X^{'}(0)-A\sin w(t+\tau)+\langle x^{'}(t)\rangle \right)
-\left(X(0)-A\sin wt+ \langle x(t)\rangle \right)\nonumber\\
&&=\left(X^{'}(0)-A\sin w(t+\tau)+\sqrt{\frac{2\hbar}{m^{\ast}w_{c}}}|\alpha_{0}|\cos w_{c}(t+\tau)\right)
-\left(X(0)-A\sin wt+\sqrt{\frac{2\hbar}{m^{\ast}w_{c}}}|\alpha_{0}|\cos w_{c}t\right)\nonumber\\
&&=\Delta X_{0}-A\left(\sin w(t+\tau)-\sin wt\right)+\sqrt{\frac{2\hbar}{m^{\ast}w_{c}}}|\alpha_{0}|\left(\cos w_{c}(t+\tau)-\cos w_{c}t\right)
\end{eqnarray}
\end{widetext}
If we consider, on average, that the scattering jump begins when the MW-driven
oscillations is at its mid-point, then $wt=2\pi n$ and being $\tau=2\pi/w_{c}$,
we end up having,
\large
\begin{equation}
\Delta X_{MW}=\Delta X_{0}-A\sin 2\pi \frac{w}{w_{c}}
\end{equation}
\normalsize

This result affects dramatically $\sigma_{xx}$ and in turn $R_{xx}$ where MIRO
are going to turn up through $\Delta X_{MW}$.
In Fig. 3 we present schematic diagrams for the different
situations regarding MIRO and ZRS.
The specific value of the evolution time and mainly the  minimum uncertainty
principle are
essential part of this discussion. These conditions fix the
final coherent state to get to after scattering.
In the dark  an electron in the initial coherent state scatters  and
jumps to the final coherent state.
On average the advanced distance is $\Delta X_{0}=X^{'}_{0}-X_{0}$.
When the light is on, depending on
the term $A\sin 2\pi \frac{w}{w_{c}}$, some times the minimum uncertainty
final state will be further  than in the dark from the initial state position.  Thus,
 on average,  $\Delta X_{MW}>\Delta X_{0}$, giving
rise to peaks  (see Fig. 3 upper panel). On the other hand, other times will be closer
 and $\Delta X_{MW}<\Delta X_{0}$ giving rise to valleys. Finally, when the driven
 coherent states are going backward, if the radiation power is
large enough, the final state will be behind the initial state in the dark (see Fig. 3 lower panel).
However, the scattered electron can only effectively jump forward due to the
DC electric field direction and the final coherent state can never be reached; in
the forward direction there is no final coherent state fulfilling the
 minimum uncertainty condition and the scattering can not be completed. Thus,
the system reaches the ZRS scenario where the electron remains in the
initial coherent state.
\newline
\subsection{Schrodinger cat states}
At this point, we can go further and apply the previous model to study MIRO in ultraclean samples were
mobility $\mu \gg 10^{7}$ $cm^{2}/Vs$. Experimental evidences\cite{zudov2,rui}
show two unexpected and striking MIRO features that stand out when dealing with
this kind of samples. First, there is an almost complete magnetoresistance collapse,
observed also in the dark and well-known as {\it giant negative magnetoresistance}\cite{bock,kriisa,inagnm}.
Second, there is a resonance peak shift from the expected position ($w_{c}=w$) to
 $2w_{c}=w$, similar
to the second harmonic generation in quantum optics. 
Following our model,
we explain the new experimental features based on one of the
fundamental  principles of quantum mechanics: the principle of  superposition of
quantum states. In our present case, the superposition of coherent states of
quantum harmonic oscillator. Paradigmatic examples of the superposition of two
coherent states are the even and odd coherent states. These states are
superpositions of two coherent states of equal amplitude but separated
in phase by $\pi$ radians:
\begin{equation}
|\alpha \rangle _{\binom{even}{odd}}=\frac{1}{2} N_{\binom{e}{o}}
\left[|\alpha \rangle \pm |-\alpha \rangle\right]
\end{equation}
where $N_{e}=\frac{e^{|\alpha|^{2}/2}}{\sqrt{\cosh(|\alpha|^{2})}}$ for even coherent states and
 $N_{o}=\frac{e^{|\alpha|^{2}/2}}{\sqrt{\sinh(|\alpha|^{2})}}$ for odd coherent states.
The even and odd coherent states can be obtained from the ground state with
the action of the even and odd displacement operators\cite{dodonov},
\begin{equation}
D(\alpha_{even})=\frac{1}{2}[D(\alpha)+D(-\alpha)]=\cosh (\alpha a^{\dagger}-\alpha^{\ast}a)
\end{equation}
and
\begin{equation}
D(\alpha_{odd})=\frac{1}{2}[D(\alpha)-D(-\alpha)]=\sinh (\alpha a^{\dagger}-\alpha^{\ast}a)
\end{equation}
 Thus,
$|\alpha \rangle _{\binom{even}{odd}}=D(\alpha_{\binom{even}{odd}})|\phi_{0}\rangle$,
and the wave function reads, $\psi_{\alpha \binom{even}{odd}}=\langle x|D(\alpha_{\binom{even}{odd}})|\phi_{0}\rangle$:
\begin{widetext}
\begin{equation}
\psi_{\alpha \binom{even}{odd}}=\frac{1}{2} N_{\binom{e}{o}}e^{i\vartheta_{\alpha}}
\left[e^{\frac{i}{\hbar}\langle p \rangle x}
\phi_{0}[x-X(0)-\langle x \rangle(t)] \pm e^{-\frac{i}{\hbar}\langle p \rangle x}
\phi_{0}[x-X(0)+\langle x \rangle(t)]\right]
\end{equation}
\end{widetext}
Thus,   in the position representation, both types of superpositions
are made up of two Gaussian wave packets oscillating back and forth periodically
at the same frequency $w_{c}$ but with a phase difference of $\pi$ (see Fig. 4).
The corresponding expansions  for the even coherent state including the time evolution read,
\begin{equation}
| \alpha(t) \rangle_{even}=\frac{e^{-iwt/2}}{\sqrt{\cosh(|\alpha|^{2})}}\sum_{n}\frac{(\alpha_{0} e^{-iw_{c}t})^{2n}}{\sqrt{(2n)!}}|\phi_{2n}\rangle
\end{equation}
and for the odd coherent state,
\begin{equation}
| \alpha(t) \rangle_{odd}=\frac{e^{-iwt/2}}{\sqrt{\sinh(|\alpha|^{2})}}\sum_{n}\frac{(\alpha_{0} e^{-iw_{c}t} )^{2n+1}}{\sqrt{(2n+1)!}}|\phi_{2n+1}\rangle
\end{equation}
Accordingly, even coherent states are a superposition of even eigenstates of the quantum harmonic oscillator and the
states energies are given by $E_{even}=\hbar w_{c}(2n+1/2)$. The odd ones
are superpositions of odd eigenstates and the energies are, $E_{odd}=\hbar w_{c}((2n+1)+1/2)$.
Remarkably enough, in both of them only one every other Landau level is populated and thus
the energy difference between populated states is $2\hbar w_{c}$ (see Fig. 4).
\begin{figure}
\centering \epsfxsize=3.5in \epsfysize=2.3in
\epsffile{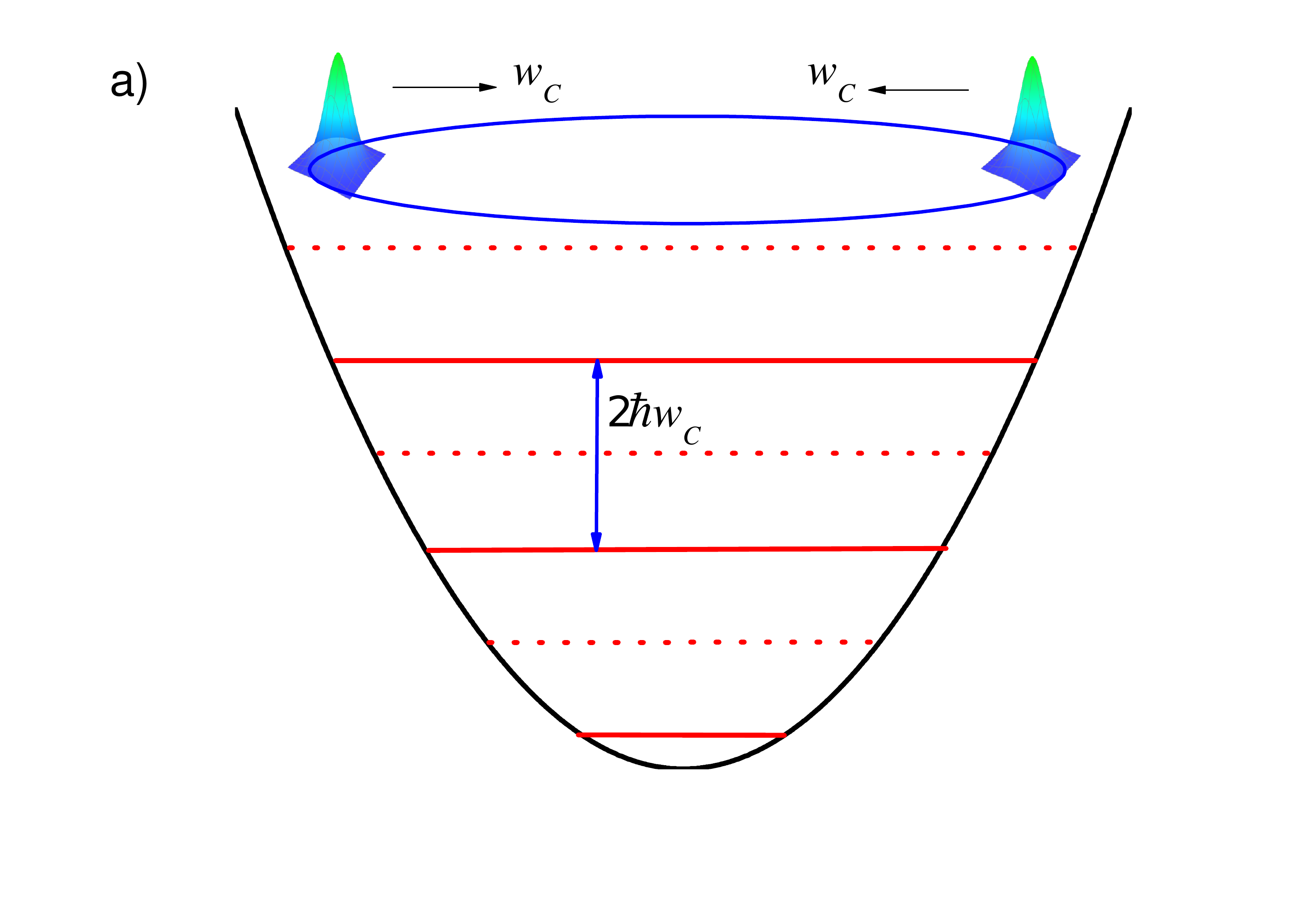}
\centering \epsfxsize=3.5in \epsfysize=2.5in
\epsffile{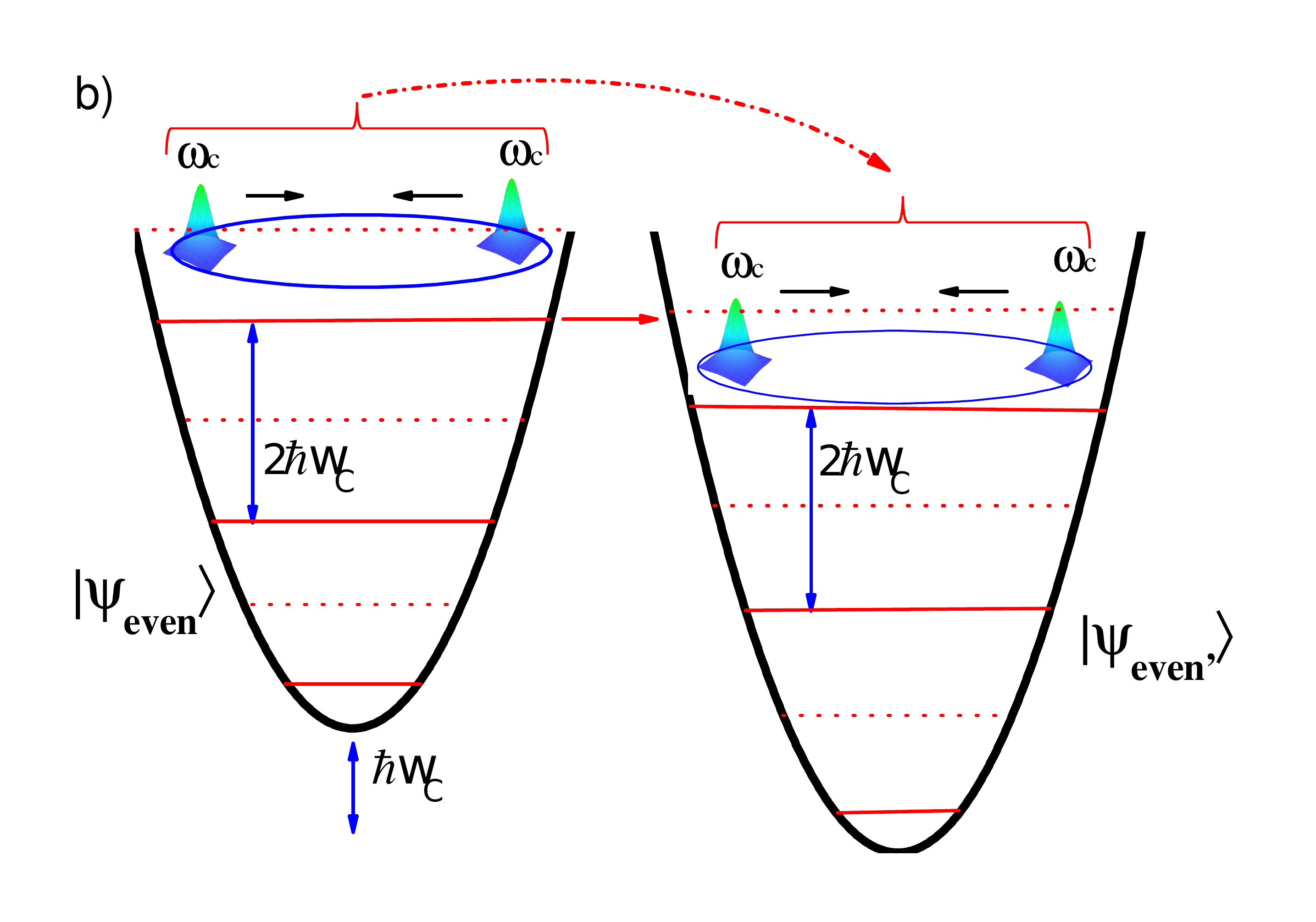}
\centering \epsfxsize=3.in \epsfysize=2.0in
\epsffile{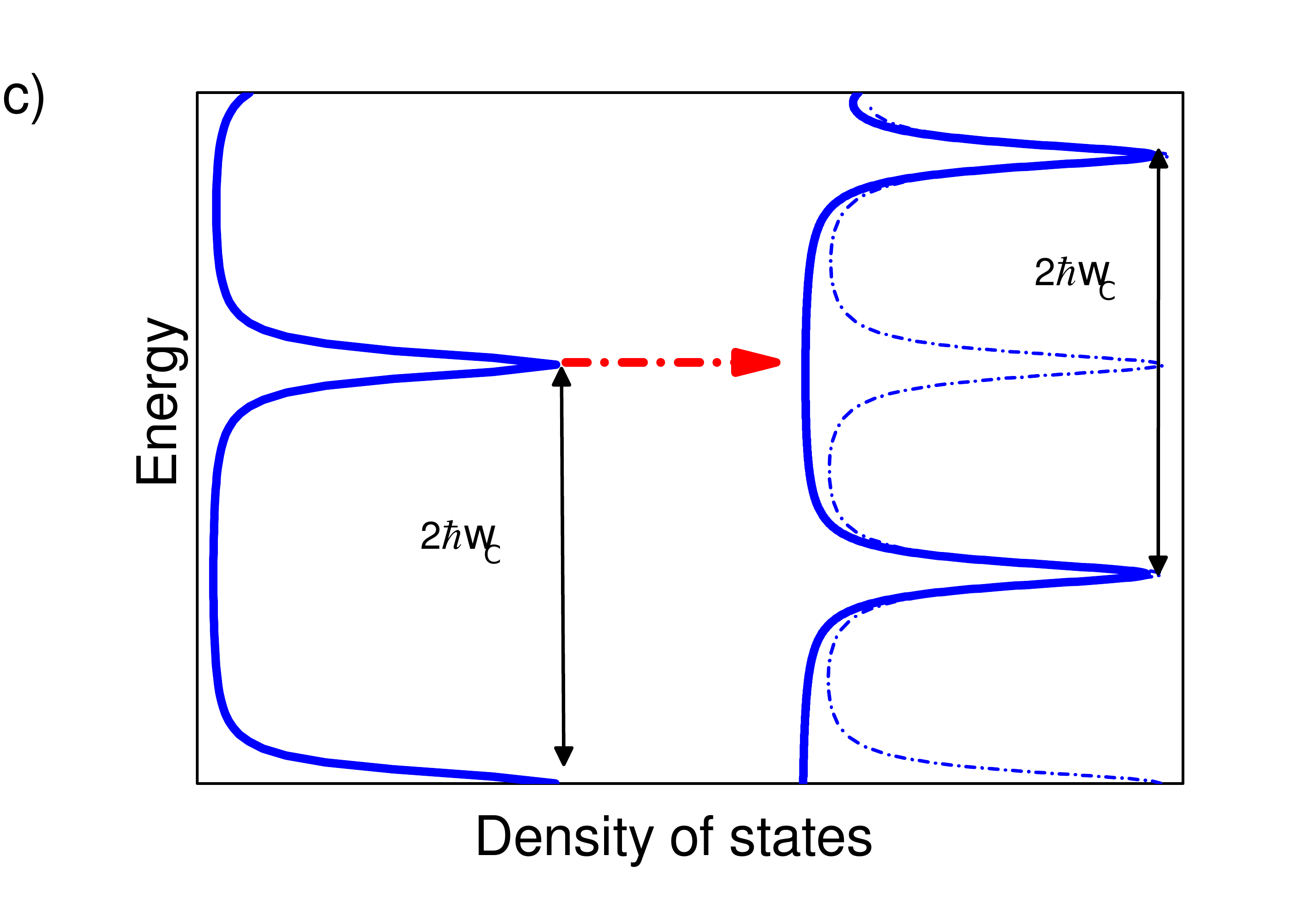}
\caption{a) Schematic diagram of Schrodinger cat states levels structure with the energy
difference of $\Delta E=2\hbar w_{c}$. b) Diagram of scattering jump
between Schrodinger cat states. Every car state is made up of two opposite
phase wave functions. c) Diagram showing the quasielastic
nature of charged impurity scattering that in these states leads to a $R_{xx}$ collapse
due to states misalignment.}
\end{figure}

When $|\alpha|$ is large, as in our case for low $B$, the coherent states $|\alpha \rangle$
and $ |-\alpha \rangle$ can be considered as macroscopically distinguishable, i.e, the
two gaussian wave packets are located at macroscopically separated points.
Then, electrons  are simultaneously  localized
in both spatially separated wave packets at macroscopic distances of the
order of the low $B$ cyclotron radius (see Fig. 4).
Thus, the above superpositions are referred as {\it Schrodinger cat states}\cite{yurke,noel} and for them
$N_{e}\simeq N_{o} \simeq \sqrt{2}$.
These cat states are mainly used in quantum optics and have recently become relevant
in quantum computing
 as a promising platform to implement qbits\cite{edit}. Now, in this new scenario, we calculate $R_{xx}$ with
 Schrodinger cat states wave functions  (Eq. (33)) and following again a semiclassical Boltzmann model (see Fig. 4)
 .
 We first calculate the scattering rate where the essential part is
the integral,
\begin{widetext}
\large
\begin{equation}
I_{\alpha,\alpha^{'}}=\int^{\infty}_{-\infty} \psi_{ \alpha^{'}\binom{e}{0}} e^{i q_{x} x}\psi_{\alpha \binom{e}{0}} dx
\propto e^{-2|\alpha_{0}|^{2}\sin^{2}(w_{c}(t+\frac{\tau}{2})-\varphi) \sin^{2}(w_{c}\frac{\tau}{2})} +
        e^{-2|\alpha_{0}|^{2}\cos^{2}(w_{c}(t+\frac{\tau}{2})-\varphi) \cos^{2}(w_{c}\frac{\tau}{2})}
\end{equation}
\normalsize
\end{widetext}
For similar reasons as above, this integral is negligible due to large $|\alpha_{0}|$. Accordingly, the scattering rate
and $R_{xx}$ are negligible too. Nonetheless,  there are important exceptions: first when the evolution time, $\tau= \frac{2\pi}{w_{c}}$, that
makes  the first exponential
 equal to one. And second, when
$\tau= \frac{2\pi}{2w_{c}}$ where the second equals one. Thus, generalizing,  when
\begin{equation}
\tau=n \times \frac{2\pi}{2w_{c}}
\end{equation}
$n$ being a positive integer.
This an important result because according to it,  the joint effect of
two gaussian wave packets oscillating at $w_{c}$ with a phase difference
of $\pi$ gives the same result as one coherent state with a wave packet oscillating
at double frequency $2w_{c}$. Besides, the energy difference between Landau
levels for both systems  is the same, $2w_{c}$, and
for large values of the Landau level index, the corresponding states energies are nearly alike.
Thus, for even cat states,
$E_{2n}=\hbar w_{c}(2n+1/2) \simeq \hbar (2 w_{c})(n+1/2)$.
Similar point applies for odd cat states.

Therefore, from the scattering point of view, the physics of
Schrodinger cat states of the quantum harmonic oscillator with a certain
frequency, $w_{c}$,  is the same as one coherent state with double frequency.
This has to be reflected in scattering-dependent physical properties.
For instance, the experimentally-obtained\cite{zudov2,rui} resonance peak shift in the MIRO's profile  that now
 shows up at twice the initial cyclotron frequency, $2w_{c}=w$.
On the other hand, this double frequency coherent state would also minimize
the Heisenberg uncertainty principle. This condition along with that
 the energy difference is, as above,  $\Delta E= \hbar w_{c}$ implies
that only  scattering processes with
$\tau= \frac{2\pi}{w_{c}}$, i.e., $n =2$, would comply with this principle.
However the ones with $n=1$, ($\tau= \frac{2\pi}{2w_{c}}$) would  not. Other processes reaching cat states at  $\Delta E= 2 \hbar w_{c}$
imply longer scattering jumps and then their contributions to the current are much smaller although
fulfilling the minimum uncertainty principle with $\tau= \frac{2\pi}{2 w_{c}}$.
Importantly enough, for quasielastic scattering processes this leads the system to a strong misalignment situation between the
scattering-involved cat states (see Fig. 4) giving rise to a dramatic collapse
of magnetoresistance, either in the dark (giant negative magnetoresistance) or under illumination.

We can generalized the above theory to Schrodinger cat states with more subcomponents.
For instance three or four coherent states that have been already experimentally carried out\cite{vlastakis}.
Thus, we can think of a {\it triangular} Schrodinger cat state with the expression:
\begin{eqnarray}
|\alpha _{3} \rangle&=&\frac{1}{\sqrt{3}}  \left[|\alpha \rangle + |e^{i2\pi/3}\alpha \rangle+|e^{i4\pi/3}\alpha \rangle \right]\nonumber\\
&=&\sqrt{3} e^{-|\alpha|^{2}/2}\sum_{n}\frac{\alpha^{3n}}{\sqrt{(3n)!}}|\phi_{3n}\rangle
\end{eqnarray}
with  an energy of $E_{3n}=\hbar w_{c}(3n+1/2)$ for the levels in $|\alpha _{3} \rangle$ and an energy difference
between them of $\Delta E_{3n}=3 \hbar w_{c}$. Therefore, one out of three levels is populated.
We can obtain the expression of the wave function similarly as the two component cat state and
thus we get to,
\begin{widetext}
\begin{equation}
\psi_{\alpha_{3}}(x,t)=\frac{1}{\sqrt{3}} e^{i\vartheta_{\alpha}}
\left[e^{\frac{i}{\hbar}\langle p \rangle(t) x}
\phi_{0}[x-X(0)-\langle x \rangle(t)] + e^{-\frac{i}{\hbar}\langle p_{1} \rangle(t) x}
\phi_{0}[x-X(0)+\langle x_{1} \rangle(t)+e^{-\frac{i}{\hbar}\langle p_{2} \rangle(t) x}
\phi_{0}[x-X(0)+\langle x_{2} \rangle(t)]\right]
\end{equation}
\end{widetext}
where $\langle x_{1} \rangle(t)$ and $\langle x_{2} \rangle(t)$ have similar expression to
$\langle x \rangle(t)$ with a phase difference of $2\pi/3$ and $4\pi/3$ respectively.
Similar conditions applies to $\langle p_{1}\rangle(t)$ and $\langle p_{2}\rangle(t)$ regarding $\langle p\rangle(t) $.
As above, we use the wave function of the {\it triangular} Schrodinger cat state in a semiclassical Boltzman model
 to calculate first the scattering rate and finally  dark
and irradiated $R_{xx}$  along  with MIRO. After similar algebra as before, we obtain that $R_{xx}$
is in general negligible except when the evolution time is given by,
\begin{equation}
\tau=n \times \frac{2\pi}{3w_{c}}
\end{equation}
Thus, again we can conclude, similarly as in the previous case,
that the joint effect of three Gaussian wave packets oscillating at $w_{c}$
with a phase difference of  $2\pi/3$ and $4\pi/3$ among them, will  produce
similar results, in terms of scattering, as only one coherent state of
triple frequency $3w_{c}$. This a prediction because the corresponding
experiments have not been carried out yet. Thus, we would expect that
$R_{xx}$ will plummet, in the dark and with light
 and remarkably, there will be
a resonance peak shift at $3w_{c}=w$. Thus, in the same way, with a
hypothetical  $n$-component
Scattering cat state we will expect a peak shift at $n\times w_{c}=w$.

\section{Results}
\begin{figure}
\centering \epsfxsize=3.6in \epsfysize=3.5in
\epsffile{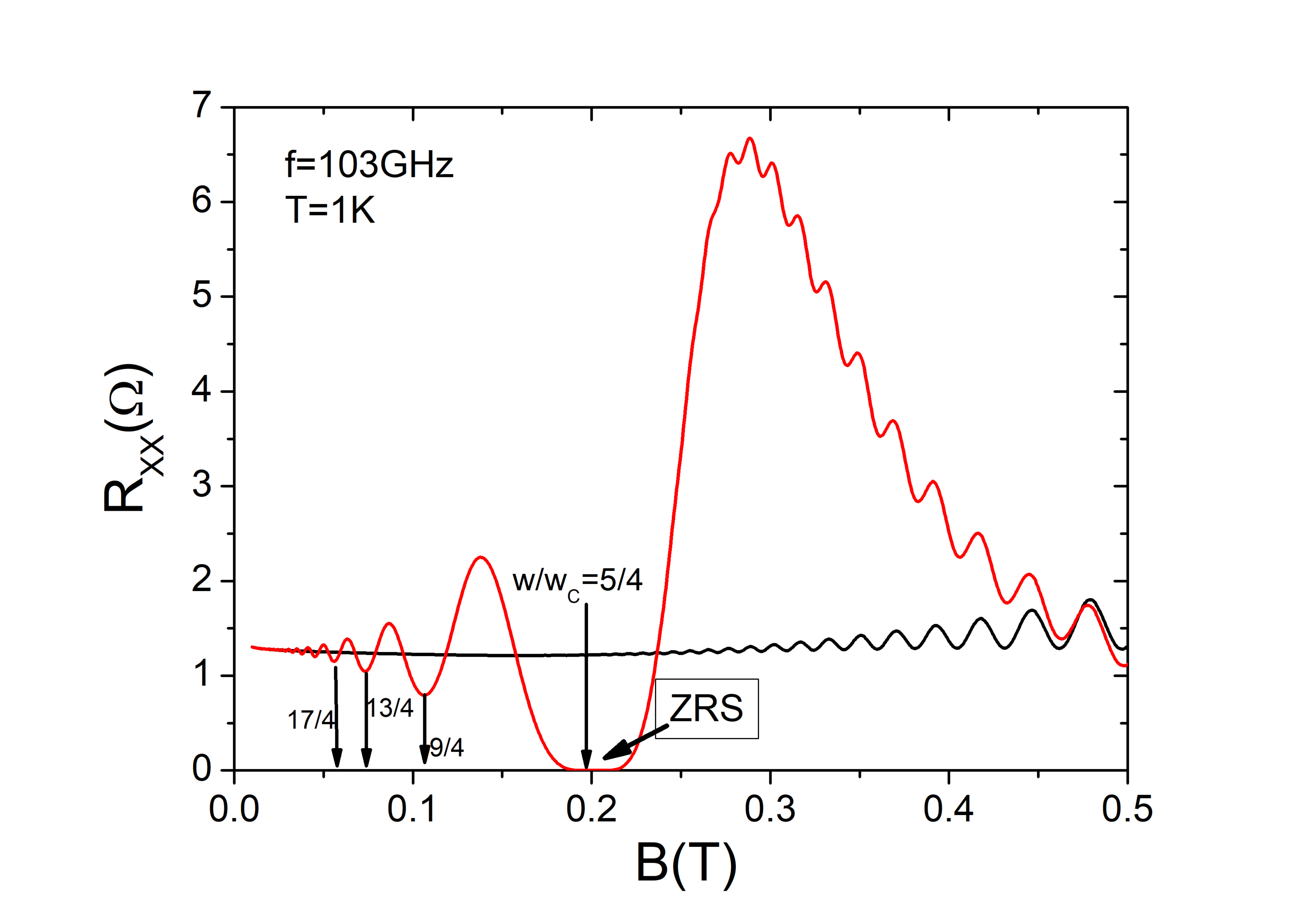}
\caption{Calculated  magnetoresistance   as a function
of $B$,  for a radiation frequency of $103$ GHz and $T=1$ K.
The dark case is also exhibited. Minima positions are indicated with arrows
corresponding to the equation,
$\frac{w}{w_{c}}=j+\frac{1}{4}$, $j$ being a positive integer.
Zero resistance states are obtained around $B\simeq 0.2T$.}
\end{figure}
\begin{figure}
\centering \epsfxsize=3.7in \epsfysize=3.in
\epsffile{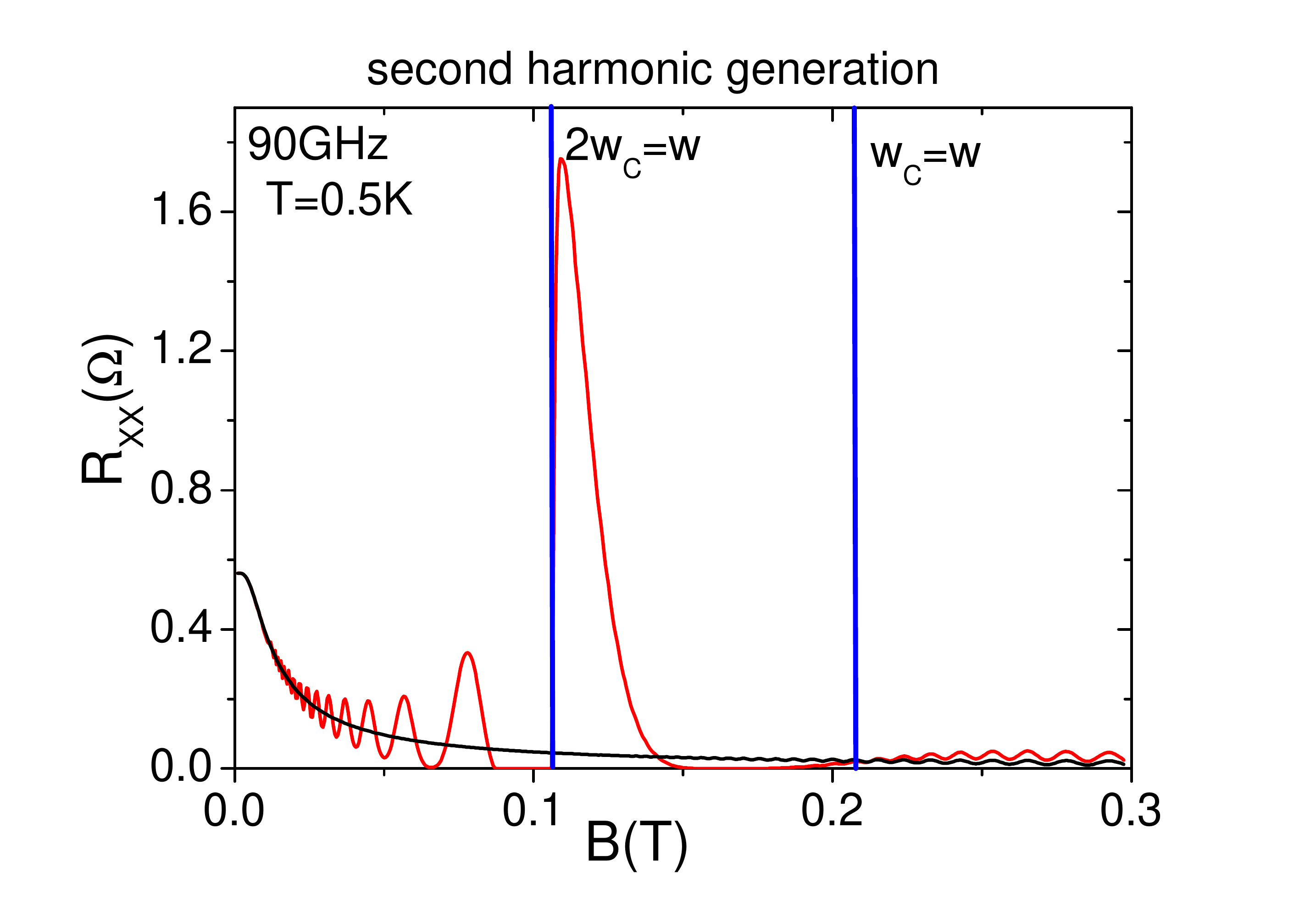}
\caption{Calculated magnetoresistance vs $B$  for a radiation frequency of $90$ GHz and $T=0.5$ K.
We also exhibit dark magnetoresistance (black curve). The
dark curve shows a dramatic  collapse or giant negative magnetoresistance
effect. When radiation is on, the curve shows MIRO, zero resistance states
and the rising of a shifted resonance peak at $2w_{c}=w$.}
\end{figure}

In Fig. 5 we present calculated results of irradiated
$R_{xx}$ vs $B$  for a radiation frequency of $103$ GHz and $T=1$ K.
The dark case is also exhibited.
In our simulations all results have been based on
experimental parameters corresponding to the experiments by Mani
\cite{mani}.
 We obtain clear MIRO where the
minima positions are indicated with arrows and, as in experiments\cite{mani},
correspond to the equation,
\begin{equation}
\frac{w}{w_{c}}=j+\frac{1}{4}
\end{equation}
$j$ being a positive integer.
Minima positions show a clear $1/4$-cycle shift, which is
a universal property that features MIRO and shows up
in any experiment about MIRO irrespective of carrier and
platform.  In the minima
corresponding to $j=1$, ZRS are found.
Now with the help of our model based on coherent states we can
explain such a peculiar value for the minima position.
Thus,  it is straightforward to check out
that the  latter equation corresponds to minima values
of MIRO.  Accordingly, $-\sin (w \frac{2\pi}{w_{c}})=-\sin(\pi/2+2\pi j)$
 would give minima values for $\sigma_{xx}$ and $R_{xx}$.
In the $sine$ argument we can identify the evolution time $\tau=2\pi/w_{c}$. Therefore,
 the value $2\pi/w_{c}$  obtained from the minima positions, would be
the "smoking gun" that would reveal that coherent states of
quantum harmonic oscillators would sustain magnetorresistance in the
dark and irradiated in high mobility 2DES.
\begin{figure}
\centering \epsfxsize=3.5in \epsfysize=3.in
\epsffile{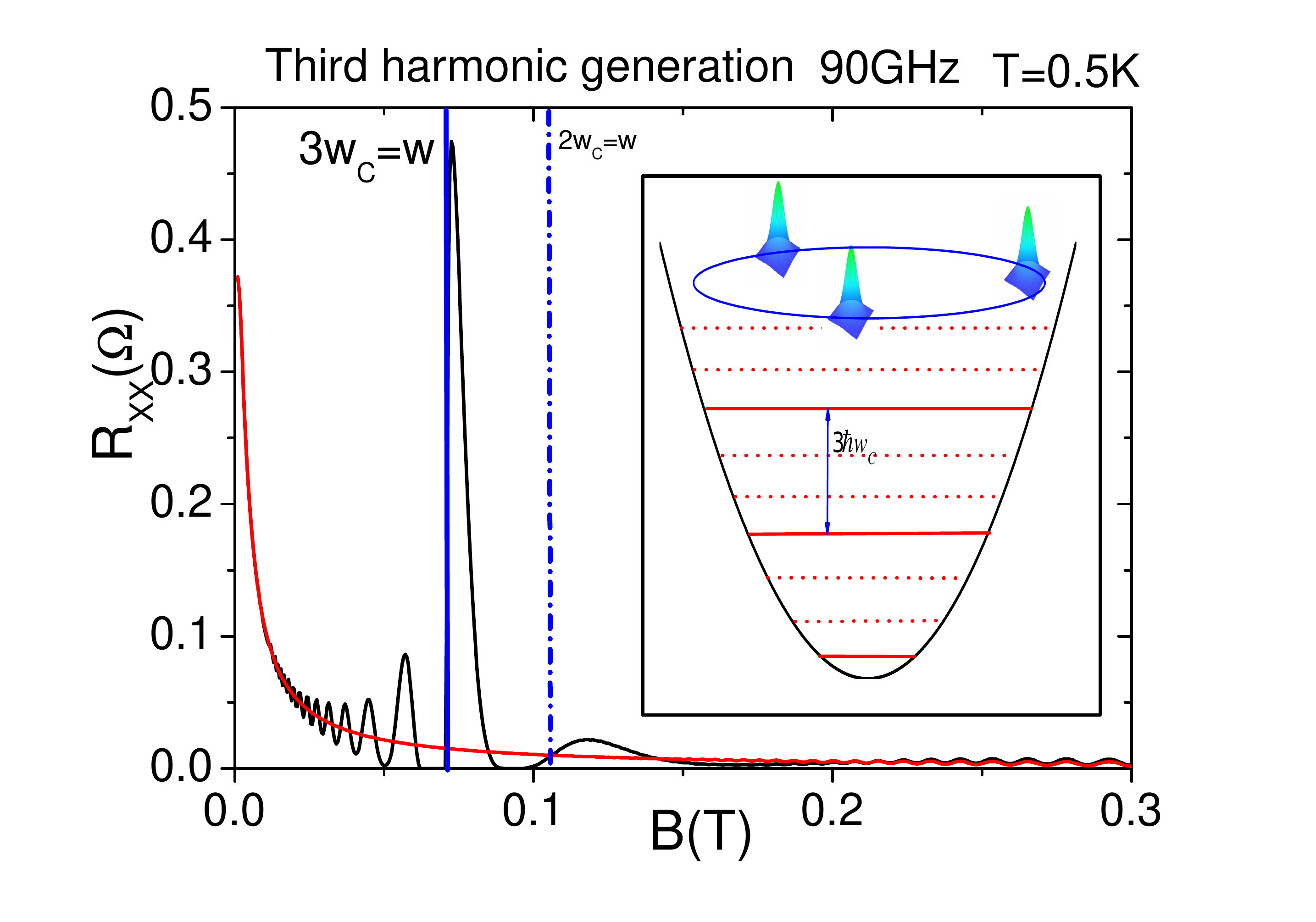}
\caption{Calculated magnetoresistance vs $B$ in the dark and with light for
a three-component Schrodinger cat state.  The radiation frequency is $90$ GHz and $T=0.5$ K.
The
dark curve shows how current plummets exhibiting  a dramatic  collapse, also known as,   giant negative magnetoresistance.
 When radiation is on, the curve shows MIRO, zero resistance states
and the rising of a shifted resonance peak at $3w_{c}=w$.}
\end{figure}

In Fig. 6, we exhibit calculated results of irradiated
$R_{xx}$ vs $B$  for a radiation frequency of $90$ GHz and $T=0.5$ K.
We exhibit also magnetoresistance without irradiation and the corresponding
dark curve shows a surprising strong $R_{xx}$ collapse or giant negative magnetoresistance
effect. When radiation is on, the curve also shows MIRO and zero resistance states, but the most
striking effect is the rise of a shifted resonance peak at $2w_{c}=w$ instead of
the expected position, $w_{c}=w$.
 All these effects above,
obtained in simulations, are in
agreement with experimental results\cite{zudov2,rui}
and can be explained with our theory.
Accordingly, magnetoresistance and MIRO are sustained by
Schrodinger cat states (even or odd) having an energy difference
between levels of $2\hbar w_{c}$.
Thus, the magnetoresistance collapse can be explained in terms of states misalignment because, as explained above,
when dealing with ultra-high clean samples the Landau levels are
so narrow that any kind of misalignment between initial and final
states gives rise to a dramatic drop of magnetoresistance. In our case,
the chaarged impurity scattering  is quasielastic
 and the energy difference between scattering-involved cat states is $\hbar w_{c}$ (see Fig. 4).
Thus, the misalignment between cat states is patent.
The resonance  peak shift is easily explained in terms of
the  new characteristic frequency of the states that now
is $2w_{c}$.  This new frequency
in the amplitude of MW-driven oscillation gives rise to the resonance peak
shift at  $2w_{c}=w$ where  is specially apparent at low $T$ and
high $P$.

In Fig. 7, we present the generalization of the model. We exhibit dark and irradiated $R_{xx}$ vs $B$ for
 three component Schrodinger cat states. Temperature and frequency are the same
 as the previous figure. A vertical single line marks the position for $3w_{c}=w$
 where a very patent resonance peak is observed. Both, dark and irradiated $R_{xx}$ curves
show how the current plummets similarly as in the two component cat state. The inset exhibits a schematic diagram with the
energy levels structure for the triangular cat state where  $\Delta E=3\hbar w_{c}$. We
would expect that increasing the number of components of the cat state, for instance {\it n-component},
the resonance peak would shift regularly to the right in $n \times \hbar w_{c}$.
\newline


\section{Conclusions}
Summing  up we have reported an extension  of the microwave-induced resistance oscillations based on
 the coherent states of the quantum harmonic oscillator. First of all, we  have obtained an expression of
the coherent states of MW-driven quantum harmonic oscillators. These MW-driven states have
been used to calculate irradiated magnetoresistance finding that
the principle of
minimum uncertainty of coherent states is crucial to understand MIRO  and zero resistance states.
Thus, different MIRO properites,
such as their physical  origin, their periodicity with the
inverse of the magnetic field, the minima and maxima position
 and the existence of zero resistance states have been explained.
Interestingly enough, for the
case of ultra-high mobility samples,  we have introduced the quantum
superposition of coherent states giving rise to Schrodinger cat states.
Based on them, we explain the experimentally obtained
magnetoresistance resonance
peak shift to  $2w_{c}=w$. This effect is similar to the one described in quantum optics
 as a second harmonic generation process. In the same way, we explained the dramatic
magnetoresistance drop that shows up in this kind of samples in terms of states misalignment.
We have generalized the model introducing three-component Schrodinger cat states and
predicted that the resonance peak will further shift to  $3w_{c}=w$ or third
harmonic generation. We have also obtained that with those states the current
will keep plummeting showing a dramatic $R_{xx}$ drop                                                                                                                                                                                             .


We acknowledge useful discussions with 
 G. Platero. 
This work was supported by the MCYT (Spain) grant PID2020-117787GB-I00.

\end{document}